\input{aipcheck}

\documentclass[
    ,final            
    ,draft            
    ,numberedheadings 
  ]
  {aipproc}

\layoutstyle{6x9}

\begin{document}

\title{\textbf{Light Cone analysis of relativistic first-order in the gradients hydrodynamics}}

\author{D. Brun-Battistini}{}
\author{A. Sandoval-Villalbazo}{
address={Departamento de F\'isica y Matem\'aticas, Universidad Iberoamericana Ciudad de M\'exico, Prolongaci\'on Paseo de la Reforma 880, Lomas de Santa Fe, M\'exico D. F. 01219, M\'exico.}
}

\begin{abstract}
 This work applies a Rayleigh-Brillouin light spectrum analysis in order to establish a causality test by means of a \textit{frequency cone}. This technique allows to identify forbidden and unforbidden regions in light scattering experiments and establishes if a set of linearized transport equations admits causal solutions. It is shown that, when studying a relativistic fluid with its acoustic modes interacting with light, Eckart's formalism yields a non causal behavior. In this case the solutions describing temperature, density and pressure fluctuations are located outside the frequency cone. In contrast,  the set of equations that arises from  modified Eckart's theory (based on relativistic kinetic theory)  yields solutions that lie within the cone, so that they are causal.
\end{abstract}

\maketitle


\section{Introduction}

  It is well-known  that Eckart's formalism contains generic instabilities while describing linear fluctuations in a relativistic fluid \cite{HL}. This fact is due to the acceleration term present in the constitutive equation for the heat flux $J^{i}_{q}$ \cite{Nos1}.
  \begin{equation}
  J^{i}_{q}=-\kappa(T^{'i}+\frac{T}{c^{2}}a^{i})
  \label{uno}
  \end{equation}
  In this paper we focus on the causality issue with a novel approach. Since light interacts with traveling waves in a fluid, absorbtion and emission processes cause frequency shifts depending on the wave speeds. The possible values of these shifts are restricted by special relativity and its description depends on linearized relativistic hydrodynamics.
In this paper we compare the predictions of the intensity spectrum of scattered light off acoustic modes as given by Eckart's formalism and the modified relativistic hydrodynamics as included in Ref. \cite{physica A}.

In section II the balance equations for a relativistic fluid are linearized using Eckart's formalism, and the corresponding spectrum equation is calculated. In section III  analog calculations are performed using the modified Eckart's formalism. In section IV the Rayleigh-Brillouin spectrum is described as the test methodology is proposed, introducing the light-frequency cone for fluid perturbations. This section also includes comparisons between the spectra arising from sections II and III. Section V is devoted to final remarks.

\section{Linearized transport equations with Eckart's formalism}
For a simple, relativistic, non-degenerate fluid we may work with the  balance equations as given below \cite{L&L} \cite{cer} \cite{Eckart}.

The continuity equation:
\begin{equation}
\frac{\partial }{\partial t}(n)+n\,\theta =0, \end{equation}
where $\theta$ is defined, as usual, as $u_{;\nu}^{\nu}=\theta$, and $n$ is the number of particles per unit of volume.

The momentum balance equation:
\begin{equation}
\tilde{\rho}\frac{\partial}{\partial t}(\theta)+n k\nabla^{2}(T)+kT \nabla^{2}( n)-A\nabla^{2}(\theta)-
\frac{\kappa}{c^{2}}\nabla^{2}\frac{\partial T}{\partial t}-\frac{\kappa T}{c^{4}} \frac{\partial^{2} \theta}{\partial t^{2}}=0,  \label{momL}
\end{equation}
where $k$ is the Boltzmann constant, $\kappa$ is the thermal conductivity and $A$ is the longitudinal kinematic viscosity \cite{berne}. The quantity $\tilde{\rho}=\frac{1}{c^{2}}(n \varepsilon + p)$ is identified with the local enthalpy of the fluid. Use has been made of the equation of state $p=nkT$.

And, finally,  the energy conservation equation:

\begin{equation}
\frac{\partial }{\partial t}(T)-\frac{2 T}{3}  \theta -D_{T}\nabla^{2}( T)- \frac{D_{T}T}{c^{2}}\frac{\partial \theta}{\partial t} =0, \end{equation}
with $D_{T}$ the thermal diffusivity \cite{berne}.

For a proper linearization of this set of equations we decompose each thermodynamic function ($T$, $n$ and $\theta$) in terms of its value at equilibrium and a fluctuation term \cite{cer}, that is
\begin{equation}
X = X_{o} + \delta X
\end{equation}
Subscripts "$o$" refer to values at equilibrium. Making this substitution and neglecting second order terms in fluctuations, the following set of linearized equations, with Eckart's formalism arises.
For particle conservation:
\begin{equation}
\frac{\partial }{\partial t}(\delta n)+n_{o}\,\delta \theta =0
\end{equation}
For the momentum balance:
\begin{equation}
\tilde{\rho}_{o}\frac{\partial (\delta \theta)}{\partial t}+ k T \nabla^{2}(\delta n) + nk \nabla^{2}(\delta T)-A\nabla^{2}(\delta \theta)-\frac{\kappa}{c^{2}}\nabla^{2}\frac{\partial \delta T}{\partial t}-\frac{\kappa T_{o}}{c^{4}} \frac{\partial^{2} \delta \theta}{\partial t^{2}}=0 \end{equation}
And the conservation of energy equation, making the substitution $D_{T}=\frac{\kappa}{m n_{o} c_{n}}$, reads:
\begin{equation}
\frac{\partial }{\partial t}(\delta T)-\frac{2 T_{o}}{3} \delta \theta -\frac{\kappa}{m n_{o} c_{n}}\nabla^{2}(\delta T)- \frac{\kappa T_{o}}{m n_{o} c_{n}c^{2}}\frac{\partial \delta \theta}{\partial t} =0  ,\end{equation}
where $c_{n}$ is the specific heat at constant particle density.

 In order to obtain the spectrum \cite{berne}, Fourier and Laplace transforms are performed with the set of equations (6-8). Then, the following dispersion equation arises (for further details see reference \cite{physica A}):

\begin{equation}
\left|\begin{array}{ccc}
s & n_{0} & 0\\
-k T_{0} q^{2} & -\frac{\kappa T_{0}}{c^{4}}s^{2}+\tilde{\rho}_{0} s + A q^{2} & \frac{\kappa}{c^{2}}q^{2}s - n_{0}k q^{2}\\ 0 & -\frac{\kappa T_{o} s}{c^{2}m n_{o} c_{n}} - \frac{2}{3} T_{o}& s + \frac{\kappa}{m n_{o} c_{n}}q^{2}\end{array}\right|=0\label{eq:8}
\end{equation}

The evaluation of the determinant leads to a quartic polynomial that reads:

\begin{equation}
d_{4}s^{4} +s^{3} + d_{2}s^{2}q^{2} + s(d_{3}q^{4}+ d_{0}q^{2}) + d_{1}q^{4}=0, \end{equation}

the coefficients $d_{4}$, $d_{3}$, $ d_{2}$, $d_{1}$ and $d_{0}$ are given by

\begin{equation}
d_{4}=-\frac{\kappa T_{0}}{c^{4}\tilde{\rho}_{o}}, \end{equation}

\begin{equation}
d_{3}=\frac{A \kappa}{\tilde{\rho}_{o}m n_{o} c_{n}}, \end{equation}

\begin{equation}
d_{2}=\frac{A}{\tilde{\rho}_{o}}+\frac{\kappa}{m n_{o} c_{n}}(1 - \frac{2T_{o}^{2}n}{c^{2}\tilde{\rho}_{o}}),
\end{equation}

\begin{equation}
d_{1}=\frac{\kappa T_{0} k}{m c_{n} \tilde{\rho}_{o}}, \end{equation}

\begin{equation}
d_{0}=\frac{c_{p}k n_{o} T_{o}}{c_{n}\tilde{\rho}_{o}}, \end{equation} where $c_{p}=\gamma c_{n}$ is the specific heat at constant pressure.

The reader may recall that the fourth root of Eq. (10) doesn't allow a finite spectrum, since the corresponding Fourier transform is divergent \cite{physica E}. Physically, Eckart's framework predicts a significant (unobserved) broadening of the central peak, directly associated to the thermal conductivity coefficient present in Eq. (11). This coefficient appears due to the heat acceleration coupling, Eq. (1) .

\section{Fluctuation transport equations with modified Eckart's formalism}
If Eq. (1) is replaced in the basic formalism with the alternative constitutive equation, directly obtained from kinetic theory \cite{physica A}:
\begin{equation}
 J^{i}_{q}=-L_{TT}T^{'i}-L_{nT} n^{'i}
\end{equation}
The linearized continuity equation (6) remains unaltered, while the linearized  momentum balance equation is now \cite{physica A}:
\begin{equation}
\rho_{o}\frac{\partial (\delta \theta)}{\partial t}+k T \nabla^{2}(\delta n) + nk \nabla^{2}(\delta T)-A\nabla^{2}(\delta \theta)-\frac{L_{TT}}{c^{2}}\nabla^{2}\frac{\partial (\delta T)}{\partial t}-\frac{L_{nT}}{c^{2}} \frac{\partial \nabla^{2} (\delta n)}{\partial t}=0 \end{equation}
and the linearized energy balance reads:
\begin{equation}
\frac{\partial }{\partial t}(\delta T)-\frac{2 T_{o}}{3} \delta \theta -\frac{ L_{TT}} {n_{o} c_{n}}\nabla^{2}(\delta T)- \frac {L_{nT}} {n_{o} c_{n}} \nabla^{2}(\delta n)=0 \end{equation}
$L_{TT}$ is an effective thermal conductivity given by
\begin{equation}
L_{TT} = \kappa (1 + \frac{\beta T_{o}}{c^{2}\rho_{o} k_{T}}), \label{LTT}
\end{equation}
where $k_{T}$ is the thermal compressibility and $\beta$ the thermal-expansion coefficient.

Defining $z=\frac{k T}{m c^{2}}$, in the case of an ideal gas, Eq. ($\ref{LTT}$) reduces to
\begin{equation}
L_{TT} \cong \kappa (1 + z)
\end{equation}
for $z<<1$. The other coefficient, $L_{nT}$, has no counterpart in the classical regime and is given by
\begin{equation}
L_{nT} =\frac{ \kappa T_{o}}{n_{o} k_{T}  c^{2}\tilde{\rho}_{o}}\label{LnT} \end{equation}
For z$<<$1, equation ($\ref{LnT}$) reduces to
\begin{equation}
L_{nT} \cong z \frac{ T_{o}}{n_{o}} \kappa
\end{equation} for an ideal gas.
As in the previous section, Fourier and Laplace transforms are taken in the whole system, so that the following dispersion equation yields:
\begin{equation}
\left|\begin{array}{ccc}
s & n_{0} & 0\\
-\frac{1}{n_{0}\kappa_{T}}q^{2}+\frac{L_{nT}}{c^{2}}sq^{2} & \tilde{\rho}_{0}s+Aq^{2} & \frac{L_{TT}}{c^{2}}q^{2}s-n_{o}T_{o}q^{2}\\
\frac{L_{nT}}{n_{0}c_{n}}q^{2} & \frac{T_{0}^2}{c_{n}} & s+\frac{L_{TT}}{n_{0}c_{n}}q^{2}\end{array}\right|=0\label{eq:8}
\end{equation}
In this case we don't have a quartic,  but a \emph{cubic polynomial} equation:
\begin{equation}
s^{3} + b_{2}s^{2}q^{2} + s(b_{3}q^{4}+ b_{1}q^{2}) + b_{4}q^{4}=0, \end{equation}
with the coefficients given by:
\begin{equation}
b_{2}=\frac{A}{\tilde{\rho}_{o}}+\frac{L_{TT}}{n_{o}c_{n}} (1 -\frac{T_{o}^{2} n_{o}}{c^{2}\tilde{\rho}_{0}}) -  \frac {n_{o} L_{nT}} {c^{2}\tilde{\rho}_{o}}, \end{equation}
\begin{equation}
b_{3}=\frac{A L_{TT}}{n_{o}c_{n}\tilde{\rho}_{o}},
\end{equation}
\begin{equation}
b_{1}=\frac{c_{P}k n_{o} T_{o}}{c_{n}\tilde{\rho}_{o}} ,
\end{equation}
\begin{equation}
b_{4}=\frac{T_{0} k}{c_{n} \tilde{\rho}_{o}}(L_{TT} - \frac{n_{o} L_{nT}}{k}) ,
\end{equation}
The Rayleigh peak's width \cite{physica A} reads now:
\begin{equation}
\Delta_{R}=\frac{b_{4}}{b_{1}}q^{2}= \frac{q^{2} }{c_{P} n_{o}}(L_{TT} - \frac{n_{o} L_{nT}}{k}),
\end{equation}
In this last expression, the pathological broadening that arises with Eckart's formalism is absent.

\section{The Rayleigh-Brillouin spectrum and the light frequency cone for fluid perturbations}

The reader may recall that in a simple relativistic fluid two transverse and three longitudinal hydrodynamics modes are found \cite{berne} \cite{boon}. The latter are at the origin of light scattering, giving rise to what it is called the Rayleigh-Brillouin spectrum, which consists of a central peak (the Rayleigh peak) and two secondary peaks (the Brillouin peaks) that appear from a Doppler effect. Denoting $\omega_{0}$  the frequency of the center of Rayleigh peak, then the two Brillouin maxima are located at $\omega_{0}\pm c_{s}q$, where $c_{s}$ is the speed of sound and $q$ is the wave number.

When waves are present in a fluid, the causality properties can be verified by making it interact with light.
If light with a frequency $\nu$ is re-emitted with a frequency $\nu' = \nu( 1 + \frac{u}{c})$, (where $u$ is the fluctuation's propagation speed), then $\nu'$ would never become greater than $2 \nu$. Thus, \textbf{a light cone on the frequency domain is generated}.

The $d_{4}$ coefficient is the clue to determine that Eckart's formalism yields a non-causal solution \cite{physica E} \cite{oviedo}, since the \emph{width of Rayleigh's peak is now enhanced} by the amount:

\begin{equation}
\Delta\cong\frac{\tilde{\rho_{o}}c^{4}}{\kappa T_{o}}
\end{equation}
The term $cq$ delimits the frequency cone so if the following relation holds
\begin{equation}
\frac{\tilde{\rho_{o}}c^{4}}{\kappa T_{o}} > cq
\end{equation}
then the  solution is out of the cone and, thus, is \textbf{non-causal}.

Following the same reasoning included in Hiscock and Lindblom's work \cite{HL} some calculations for water at a temperature of 293 K and pressure of 1 bar are shown below:

\begin{equation}
\frac{\tilde{\rho_{o}}c^{4}}{\kappa T_{o}} \sim 10^{34}\frac{1}{sec} \end{equation}
evaluating $cq$ with the typical value $q=10^{7}m^{-1}$

\begin{equation}
cq \sim 10^{15}\frac{1}{sec}
\end{equation}

Since $10^{34} > >10^{15}$ the solution is clearly out of the frequency cone,  and thus is non-causal. Physically, this would mean that the interacting particles of the fluid move with a supraluminical speed after collisional processes.

\section{Final remarks}
A test based on light scattering experiments has been developed to determine if a set of linearized hydrodynamic transport equations yields to a causal solution or not. When performing the calculations under Eckart's formalism, and increasing $z$, the Rayleigh peak's width invade the two Brillouin ones, entering the forbidden region of the light cone. It is clear that the non-causality issue arises when coupling the heat flux with the acceleration. This work has been developed with first order in the gradients theories. With the standard version of extended theories \cite{jou}, the fourth root ($s^{4}$) is conserved, but damping terms are included. It is not clear if this damping preserves the overlapping of the Brillouin peaks mentioned above. Also, the damping coefficients cannot be kinetically supported by a Chapman-Enskog expansion. These issues deserve a closer look and will be studied in the near future.

The authors wish to thank A.L. Garcia-Perciante and L.S. Garcia-Colin for their valuable and encouraging comments.




\bibliographystyle{aipproc}   




\end{document}